\documentclass[useAMS,usenatbib,usegraphicx]{mn2e}

\usepackage{natbib,aas_macros,amsmath,amssymb}
\usepackage{color}

\title[Coronal Loop]{Competition between shock and turbulent heating in coronal loop system}
\author[Takuma Matsumoto]{Takuma Matsumoto \thanks{E-mail:mtakuma@solar.isas.jaxa.jp}\\
Institute of Space and Astronautical Science, Japan Aerospace Exploration Agency, 3-1-1 Yoshinodai,\\
 Chuo-ku, Sagamihara, Kanagawa 252-5210, Japan}
\begin{document}

\date{Accepted 1988 December 15. Received 1988 December 14; in original form 1988 October 11}

\pagerange{\pageref{firstpage}--\pageref{lastpage}} \pubyear{2002}

\maketitle

\label{firstpage}

\begin{abstract}
2.5-dimensional magnetohydrodynamic (MHD) simulations are performed with high spatial resolution 
in order to distinguish between competing models of the coronal heating problem.
A single coronal loop powered by Alfv\'{e}n waves excited in the photosphere is the target of the present study.
The coronal structure is reproduced in our simulations as a natural consequence of the transportation and dissipation of Alfv\'{e}n waves.
Further, the coronal structure is maintained as the spatial resolution is changed from 25 to 3 km, although
the temperature at the loop top increases with the spatial resolution.
The heating mechanisms change gradually across the magnetic canopy at a height of 4 Mm.
Below the magnetic canopy, both the shock and the MHD turbulence are dominant heating processes.
Above the magnetic canopy, the shock heating rate reduces to less than 10 \% of the total heating rate 
while the MHD turbulence provides significant energy to balance the radiative cooling and thermal conduction loss or gain.
The importance of compressibility shown in the present study would significantly impact on the prospects of successful 
MHD turbulence theory in the solar chromosphere.
\end{abstract}


\section{Introduction}

\begin{table*}
 \centering
 \begin{minipage}{140mm}
  \caption{Properties of the runs described in this paper. 
           The first and the second column give the model label and the grid numbers in the $x$ and $z$ direction.
           $\Delta x_{min}$ [km], $\Delta x_{max}$ [km], and $\Delta z$ [km] indicate the grid size of each run.
           $T$ [MK] and $n$ [10$^8$ cm$^{-3}$] correspond to the temperature and the number density at the loop top averaged over 
           $x\in[49.5,50.5]$ Mm, $z\in[-1.5,1.5]$ Mm, and $t\in[2,3]$ h, respectively.
           $L$ indicates the average length of the coronal portion ($T\ge0.1$ MK).}
  \begin{tabular}{@{}cccccccc@{}}
   \hline
   \hline
   Model&Resolution&$\Delta x_{min}$ [km]&$\Delta x_{max}$ [km]&$\Delta z$ [km] &T [MK]&n [10$^8$ cm$^{-3}$] & $L$ [Mm]\\
   \hline
   Run 1&(2048,128)&25&93.4&23.4&0.86&0.94&87 \\
   Run 2&(4096,256)&12.5&46.7&11.7&0.95&1.28&86 \\
   Run 3&(8192,512)&6.3&23.4&5.9&1.01&1.53&86 \\
   Run 4&(16384,1024)&3.0&11.4&2.9&1.03&1.63&84 \\
   \hline 
  \end{tabular}
  \label{tab:runs}
  \end{minipage}
\end{table*}

The coronal heating problem has been one of the primary challenges in the field of solar physics 
since \cite{1943ZA.....22...30E} first discovered the extremely hot corona above the cool photosphere. 
It is widely accepted that the ultimate energy source of the coronal heating is the solar convective motion
\citep{2006SoPh..234...41K,2009LRSP....6....3C,2012RSPTA.370.3217P,2015RSPTA.37340269D}.
However, the transportation and dissipation mechanisms acting in the corona have not yet been identified. 
As the coronal heating mechanisms are closely linked to the process of mass loss from stellar objects, 
investigation of the solar corona is quite important as regards the further advancement of astrophysics.

The primary objective of this paper is to distinguish between the competing coronal-heating-mechanism theories in the framework of wave heating models. In plasma with a high magnetic Reynolds number, like the solar corona, heating events occur on an extremely small scale in the form of
shocks \citep{1961ApJ...134..347O}, resonant absorption \citep{1978ApJ...226..650I}, phase mixing \citep{1983A&A...117..220H},
or turbulence \citep{1999ApJ...523L..93M}.
The inhomogeneity of the solar atmosphere allows this wide variety of heating mechanisms to act,
which increases the complexity of the coronal heating problem.

Of the various wave heating mechanisms, the present study focuses on the competition between shock heating and turbulent heating.
Super-radially expanding flux tubes extending from the photosphere allow Alfv\'{e}n waves
to increase the wave nonlinearity as they propagate outwardly.
The nonlinear Alfv\'{e}n waves are known to create slow and/or fast shocks through nonlinear mode conversion \citep{1982SoPh...75...35H}.
If the amplitude of the Alfv\'{e}n waves excited at the photosphere exceeds 1 km s$^{-1}$,
these waves are considered to drive the spicules \citep{1999ApJ...514..493K,2010ApJ...710.1857M},
to heat the corona \citep{2004ApJ...601L.107M,2008ApJ...688..669A}, and to accelerate the solar wind
\citep{2005ApJ...632L..49S,2006JGRA..11106101S,2012ApJ...749....8M,2014MNRAS.440..971M}.
Turbulent heating is also a plausible heating mechanism, and nonlinear interactions between the Alfv\'{e}n waves may drive magnetohydrodynamic (MHD)
 turbulence in the corona and the solar wind \citep{1999ApJ...523L..93M}.
Note that MHD turbulence is often described using the so-called reduced MHD (RMHD) formulation \citep{1990JGR....9510291Z}, and
some recent studies have applied the RMHD framework to the coronal loops 
\citep{2007A&A...469..347B,2011ApJ...736....3V,2012A&A...538A..70V}.
Recent numerical simulations suggest that shock heating is dominant below the transition region,
while turbulent heating is dominant above the transition region \citep{2007ApJS..171..520C,2014MNRAS.440..971M}.
However, the one-dimensional (1D) MHD and RMHD formulations cannot determine the most applicable heating mechanisms among
the various competing theories, such as those involving shocks and turbulence. Previously, \cite{2007ApJS..171..520C}
have investigated the competition between the shock and turbulent heating using a time-steady MHD model
of the solar wind. Their model includes phenomenological heating mechanisms involving shock and turbulence,
which should be confirmed by dynamical simulations. \cite{2014MNRAS.440..971M} have also suggested 
heating mechanism transitions from compressible to incompressible heating, using 2.5-dimensional (2.5D) MHD simulations. However,
the spatial resolution of those simulations is too low to allow turbulent structures to be resolved, generating the suspicion that a higher-resolution simulation could change the results significantly.
This situation motivates us to perform MHD simulations with high resolution in order to investigate the heating mechanisms in the coronal loops.
As coronal loops have a shorter Alfv\'{e}n transit time and smaller spatial scale than those in the open flux region,
the numerical-computation cost can be reduced significantly.

In this paper, a high-resolution MHD simulation of the coronal loops are performed under the hypothesis that
the heating rate can be maintained with higher resolution through development of the MHD turbulence.
Accordingly, the spatial resolution is changed from 25 to 3 km and, as a result, the higher resolution
leads to the formation of thinner current sheets that maintain the heating rate. 
Note that the conditions that maintain the turbulence or their three-dimensional 
(3D) extensions have not yet been determined; this is a topic of investigation for future papers.

\section{Models and Assumptions}

In this study, 2.5D MHD simulations were performed in order to mimic a single coronal loop.
The loop was assumed to be 100 Mm in length ($\equiv L$) and 3 Mm in width ($\equiv W$).
For simplicity, the curvature of the loop was ignored; thus, a straight idealized loop 
in a rectangular region ($x\in[0,100]$ Mm, $z\in[-1.5,1.5]$ Mm) was considered.
The $x$ and $z$ coordinates were assigned to the length direction along the loop axis and that across the loop, respectively.
The $y$ coordinates were assigned to the direction perpendicular to $x$--$z$ plane.
Then the compressible MHD equations for Cartesian geometry were solved:

\begin{equation}
 \frac{\partial \rho}{\partial t}+\bmath{\nabla \cdot} (\rho \bmath{v})=0,
\end{equation}

\begin{equation}
 \frac{\partial \rho \bmath{v}}{\partial t} + \bmath{\nabla \cdot} \left(
 p+\frac{B^2}{2} + \rho \bmath{v}\bmath{v}-\bmath{BB}
 \right)=\rho \bmath{g},
\end{equation}

\begin{equation}
 \frac{\partial \bmath{B}}{\partial t} = \bmath{\nabla}\times \left( \bmath{v}\times\bmath{B}\right),
\end{equation}

\begin{eqnarray}
 \frac{\partial {\cal E}}{\partial t} &+& \bmath{\nabla\cdot}\left[ \left( {\cal E}+p+\frac{B^2}{2}\right)\bmath{v}
 -\bmath{(B\cdot v) B}\right] \nonumber \\
 &=&\rho \mathbf{v \cdot g}+\bmath{\nabla \cdot} \left( \kappa \bmath{\nabla}T \right) + Q_{\rm rad},
\end{eqnarray}

\begin{equation}
 {\cal E}=\frac{p}{\gamma-1}+\frac{\rho v^2}{2}+ \frac{B^2}{2},
\end{equation}
where $\rho$, $\mathbf{v}$, $p$, $\mathbf{B}$, ${\cal E}$, and $T$ are the mass density, 
the fluid velocity, the gas pressure, the magnetic field normalized against $\sqrt{4\pi}$, 
the total energy density, and the temperature, respectively.
$\kappa$ is the Spitzer-type thermal conductivity tensor and $Q_{\rm rad}$ is the radiative cooling function. 
The mean molecular weight was assumed to be dependent on the temperature only, in order to mimic hydrogen ionization. 
More detailed descriptions of the thermal conduction,
radiative cooling, and equation of state can be found in \cite{2014MNRAS.440..971M}.
For the gravitational acceleration force, a half circle loop was assumed
\begin{equation}
 \bmath{g} = - \frac{GM_{\odot}\cos{\theta}}{\left(R_{\odot}+h \right)^2} \hat{\bmath{x}},
\end{equation}
where $r=L/\pi$, $\theta=x/r$, $h=r\sin{\theta}$, and $\hat{\bmath{x}}$ is the unit vector in the $x$ direction.

As the initial conditions, a static and isothermal atmosphere with a temperature of 10$^{4}$ K along the entire loop was established.
Hydrostatic equilibrium was assumed below 10 Mm with a bottom density of $10^{-7}$ g cm$^{-3}$.
The density distribution above 10 Mm was assumed to be proportional to $h^{^{-2}}$, which was not in the initial dynamical equilibrium state.
These initial conditions were chosen because of the numerical tractability in the initial phase, which is not a focus of
interest in the present study.
Note that the results presented in this paper correspond to a significantly later phase and do not depend on the initial conditions.
The potential field was chosen as the initial magnetic field that could be extrapolated from
the photospheric boundary conditions at $x=0$ and $100$ Mm
\begin{equation}
 B_{x}=B_0 \exp{\left[ - \left( \frac{z}{w_B} \right)^2\right]} - \frac{B_0 w_B \sqrt{\pi}}{W} {\rm erf}{\left( \frac{W}{2w_B}\right)} + B_{\rm c},
\end{equation}
where $B_0=1800$ G, $B_c=10$ G, and $w_B=0.5$ Mm.
The maximum field strength at the boundary is 1278 $G$, which corresponds to a expansion factor ($B_x(x=0,z=0)/B_x(x=L/2,z=0)$) of approximately 128.
The initial potential field was extrapolated using the vector potential rather than the scalar potential, in order to reduce the
error in $\nabla \cdot \mathbf{B}$ to round-off errors.

For the photospheric boundary, a prescribed velocity perturbation was substituted in the $y$ direction uniformly along the $z$-direction.
A random noise was assumed with a total power of $2.2$ km s$^{-1}$ within a finite frequency range ($\nu\in[2.5\times10^{-4},2\times10^{-2}]$ Hz).
Note that the total power used in this simulation is almost equivalent to the maximum amplitude of the observed photospheric velocity fluctuation,
which has been estimated to be a few kilometres per second based on the bright-point motion 
\citep{1994A&A...283..232M,1996ApJ...463..365B,2012ApJ...752...48C} or using local correlation tracking \citep{2010ApJ...716L..19M}.

Four sets of numerical simulations were performed with the same initial and boundary conditions by changing the numerical resolution.
In our previous study, a very low resolution (grid number in the $z$ direction, $N_z$ = 32 or $\Delta z = 100$ km) was employed to extend the numerical domain to
the solar-wind acceleration region \citep{2012ApJ...749....8M,2014MNRAS.440..971M}.
In this study, the target was switched from the solar wind to the coronal loop, which has a smaller system size
and a shorter relaxation time scale.
A uniform grid width was implemented in the $z$ direction, while a nonuniform grid was implemented in the $x$ direction.

The numerical scheme adopted in our simulation was the HLLD scheme \citep{2005JCoPh.208..315M}.
The second-order accuracy in both space and time were determined using the MUSCL interpolation with the minmod limiter and
Runge-Kutta integration. Finally, the flux-CT method \citep{2000JCoPh.161..605T,2005JCoPh.205..509G}
was implemented to reduce the numerical error in $\nabla \cdot \mathbf{B}$ to the round-off error.

\section{Results}

\begin{figure}
  \includegraphics[scale=1.0]{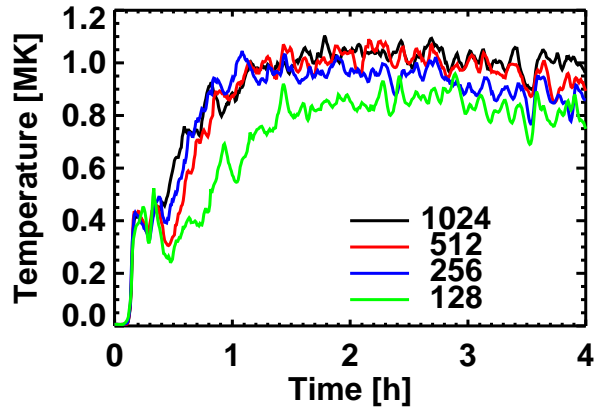}
  \caption{Temporal evolution of loop top temperature for different runs averaged over $x\in[49.5,50.5]$ Mm and $z\in[-1.5,1.5]$ Mm.
           The green, blue, red, and black solid lines indicate the Run 1--4 results, respectively.}
  \label{fig:te_evolve}
\end{figure}

The results corresponding to the various resolutions are summarized briefly in Table \ref{tab:runs}.
The model atmosphere relaxed to a quasi-steady state having a high-temperature corona 
$\sim$1.5 hr from the simulation start time (Fig. \ref{fig:te_evolve}).
For each run, the loop top temperature averaged over $x\in[49.5,50.5]$ Mm and $z\in[-1.5,1.5]$ Mm increased with time
and reached saturated states, which shall be called the quasi-steady state hereafter, although significant fluctuations
(a few percent of the mean value) over time remained.
The quasi-steady state was achieved when the radiative loss and conductive gain/loss balanced
with the heating because of the dissipation of the Alfv\'{e}n waves.
The loop top temperature, density, and loop length for Run 4 were 1.03 MK, 1.63 $\times$ 10$^8$ cm$^{-3}$, and 84 Mm on average, respectively.
Hereafter, all results are for Run 4, unless some other, specific description is provided.
The loop temperature can be predicted using the Rosner-Tucker-Vaiana (RTV) scaling law \citep{1978ApJ...220..643R}
of the coronal loop temperature as a function of the loop density and length.
The temperature predicted from the RTV scaling law, $T_{\rm RTV}=1.4\times10^3 (PL)^{1/3}$, 
where $P$ is gas pressure of the loop, is 1.0 MK, which agrees quite well with the results of our numerical simulation.

\begin{figure}
  \includegraphics[scale=1.0]{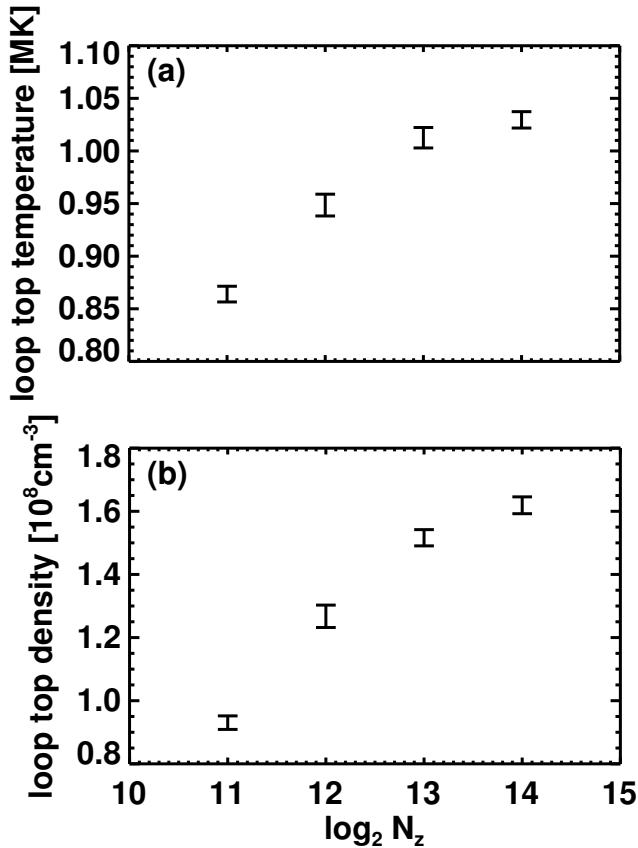}
  \caption{(a) Temperature and (b) number density as functions of grid number in $z$ direction, averaged over $x\in[49.5,50.5]$ Mm, $z\in[-1.5,1.5]$ Mm,
           and $t\in[2,3]$ h. The error bar indicates 66 \% confidence intervals for each variable.}
  \label{fig:resolve}
\end{figure}

We find that the temperature and density in the quasi-steady state depend on the spatial resolution of the numerical simulation.
Fig. \ref{fig:resolve}a shows the loop top temperature as a function of $N_z$.
The temperature increases monotonically with $N_z$ from 0.86 up to 1.03 MK.
Although the rate of increase goes down between Runs 3 and 4 (2\% difference), the difference remains statistically significant.
Therefore, we concluded that our numerical simulation had not converged at this stage.
The loop top density with respect to $N_z$ is plotted in Fig. \ref{fig:resolve}b.
The density also increases with $N_z$, which is consistent with RTV theory.
For both the temperature and density, the 66 \% confidence intervals were plotted by assuming that the time-series data obey the auto-regressive model \citep{1981ApJS...45....1S}.

\subsection{Mean profiles along loop axis }

\begin{figure}
  \includegraphics[scale=1.0]{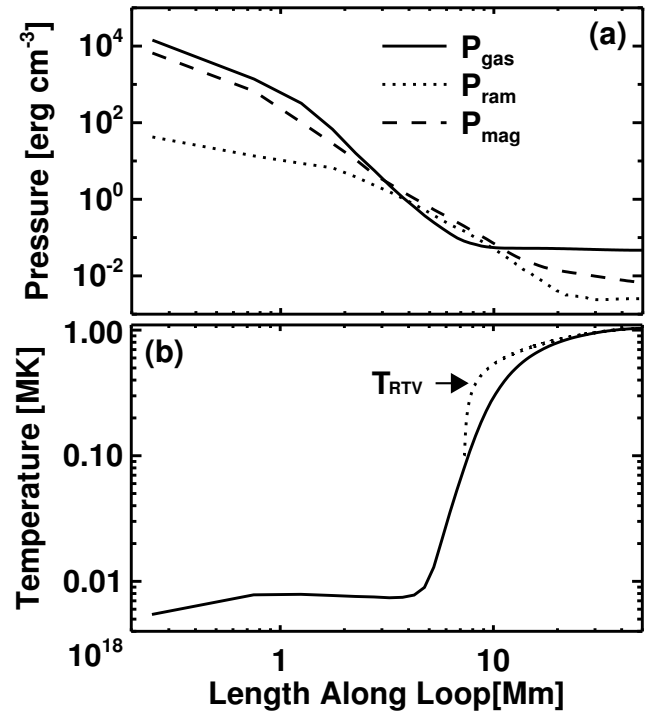}
  \caption{Mean profiles along loop axis, averaged over every 1 Mm in $x$-direction, $z\in[-1.5,1.5]$ Mm, and $t\in[2,3]$ h.
           (a) Pressure and (b) temperature of Run 4.
           The solid, dotted, and dashed lines in panel (a) correspond to the gas, ram, and magnetic pressures, respectively.
           The expected temperature profile from RTV theory is superimposed as a dotted line in panel (b).}
  \label{fig:spprof}
\end{figure}

Fig. \ref{fig:spprof}a shows the gas, ram, and transverse magnetic pressures 
($P_{gas}\equiv p$, $P_{ram}\equiv \rho v_x^2$, and $P_{mag}\equiv (B_y^2+B_z^2)/2$, respectively) 
as functions of the length along the loop.
All the variables are averaged over every 1 Mm in the $x$ direction, $z\in[-1.5,1.5]$ Mm, and $t\in[2,3]$ h.
Only the first half of the loop was plotted, but the properties described below are almost identical in the other half of the loop.
The scale height of $P_{gas}$ is approximately 200 km below 2 Mm, and increases with height to more than 200 Mm in the corona.
$P_{ram}$ exceeds $P_{gas}$ between 4 and 10 Mm, and plays an important role in the gravitational stratification.
This dynamical pressure is produced by the magnetoacoustic waves converted from the Alfv\'{e}n waves,
which is also important for the dynamical motion of the transition region.
$P_{mag}$ also exceeds $P_{gas}$, in this case, between 3 and 10 Mm.
Note that the $P_{mag}$ here primarily stems from the wave pressure of the Alfv\'{e}n waves.

The temperature profiles are shown in Fig. \ref{fig:spprof}b.
The temperature increases monotonically with height even below 1 Mm. This result differs from the standard model \citep{1981ApJS...45..635V} and this discrepancy may be due to the empirical cooling function \citep{1989ApJ...346.1010A} adopted in our model.
The height of the transition region in our model was 7.7 Mm, assuming that the transition region began at the layer with 0.1 MK.
Further, the maximum temperature in Run 4 was 1.03 MK.
The dotted line in Fig. \ref{fig:spprof}b is the temperature profile predicted by RTV theory \citep{1978ApJ...220..643R}.
When applying this theory, the loop length (84 Mm), the loop top temperature (1.03 MK), and the temperature at the coronal
bottom (0.1 MK) were specified.
The scale height of the heating function was set to infinity and the theoretical profiles were calculated using the method described in \cite{2001ApJ...550.1036A}.
In Fig. \ref{fig:spprof}b, the numerical temperature profile is similar to the theoretical profile above 20 Mm; however, a more gradual change is apparent in the former than the latter below 20 Mm.
This difference may be attributable to the nonuniform nature of the resultant heating function, which is discussed below.

\begin{figure}
  \includegraphics[scale=1.0]{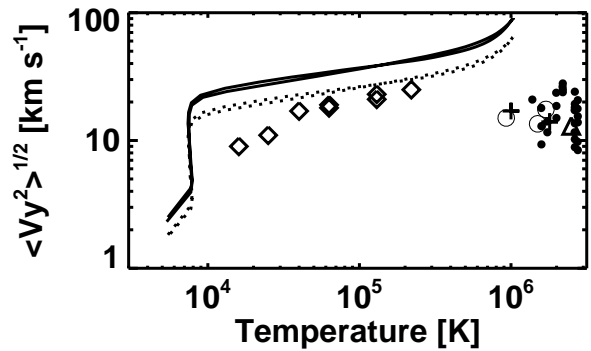}
  \caption{Root mean square of the transverse velocity as function of temperature. The solid and dotted lines correspond to
           $\langle V_y^2 \rangle ^{1/2}$ and $\langle V_y^2+V_z^2 \rangle ^{1/2}$, respectively.
           The symbols represent observational values: diamonds \citep{1978ApJ...226..698M},
           open circles \citep{1979ApJ...227.1037C}, crosses \citep{1999ApJ...513..969H},
           triangles \citep{2009ApJ...705L.208I}, and filled circles \citep{2016ApJ...820...63B}.}
  \label{fig:vy_vs_te}
\end{figure}

The root mean square of the transverse velocity ($V_y,~V_z$) is larger than the observations estimated
from the nonthermal broadening of the lines (Fig. \ref{fig:vy_vs_te}).
The transverse velocity ($V$) monotonically increases with temperature, with a sudden jump in the vicinity of 10$^4$ K due to the density stratification
in the transition region.
The $V$ of $\sim$100 km s$^{-1}$ at the loop top obtained in the present study is almost five times larger than the observational results.

\begin{figure*}[b]
  \includegraphics[scale=0.9]{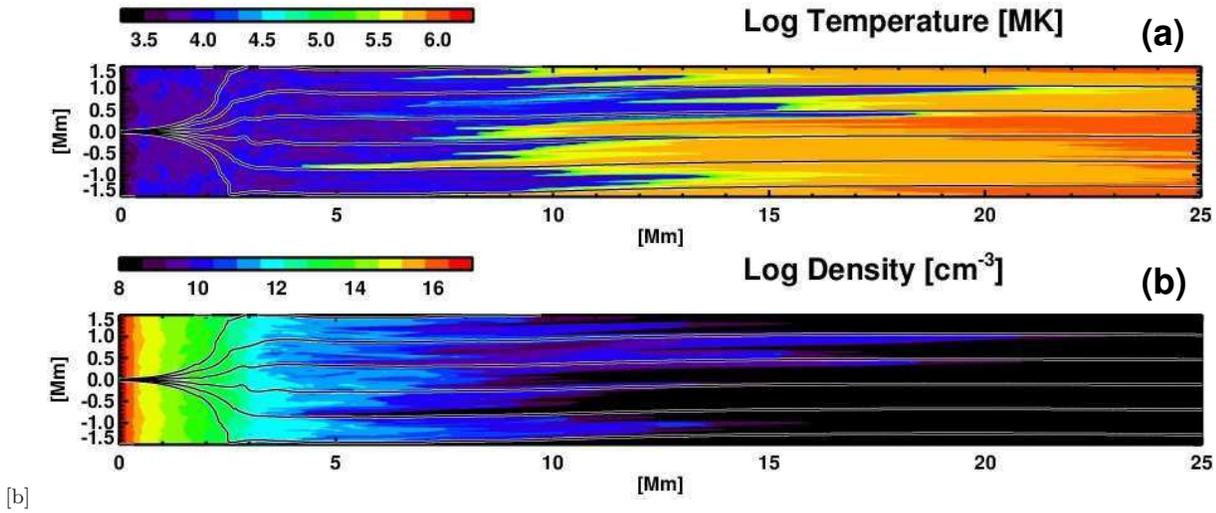}
  \caption{Snapshot of (a) temperature and (b) number density at $t=$ 2.6 h.
           The black solid lines within white envelopes indicate the magnetic field lines.
           Videos available online: Movies 1 and 2 show the temperature and number density behaviour, respectively.}
  \label{fig:snapshot_summary}
\end{figure*}

\subsection{Dynamic properties of atmosphere}

The propagation of Alfv\'{e}n waves excited at the photosphere indicates a wide variety of transient properties in the atmosphere above.
In particular, the dynamics of the model atmosphere has four remarkable features associated with Alfv\'{e}n waves (Fig. \ref{fig:snapshot_summary};
video available online: Movies 1 and 2).

First, the transition region exhibits fluctuations, which originate from the collisions between the shocks and the transition region; these collisions elevate the chromospheric materials to the coronal height \citep{1982SoPh...75...35H,1999ApJ...514..493K}.
This ascending motion is then followed by a descending motion due to the gravitational force, which can be interpreted as a spicule.

Second, the transition region is corrugated by Alfv\'{e}n waves such that it contains many peaks and valleys.
These structures are produced by the chromospheric turbulence driven by the nonlinear interaction of the Alfv\'{e}n waves.
The Alfv\'{e}n wave fronts are also corrugated before their collision with the transition region.
This corrugation feature is a distinct property of the 2D simulation and cannot be produced in 1D simulations.
Note that the corrugation process in the chromosphere is not driven by the corrugation instability \citep{1995ApJ...454..182S},
as this process applies even in the case of fast shocks, which are stable against corrugation instability.
Instead, nonlinear interactions between the magnetoacoustic waves and Alfv\'{e}n waves create fluctuations on a smaller spatial scale.
Note that the corrugation pattern can be the same phenomena as horizontally propagating surface waves excited by photospheric perturbation
found by \cite{2011ApJ...727...17F}.

Third, the resultant loop consists of numerous thin threads having an almost isothermal nature and no clear typical width.
The differential emission measure constructed at the loop top has a value of 3.3 $\times$ 10$^{19}$ cm$^{-5}$ K$^{-1}$ at
its peak, with a narrow distribution (0.13 in $\log_{10} T$ K).
The power spectra of the density, temperature, and synthesized intensity of AIA 171$\AA$ in the $z$ direction show a power law distribution
with a cut-off in the vicinity of the dissipation scale ($\sim$100 km for Run 4).
There is no elemental scale other than the dissipation scale and the smaller-scale length appears in the simulation with the higher-resolution run.

Finally, the peaks and valleys in the transition region exhibit gentle oscillations. These oscillations correspond to the manifestation of compressible fast waves in the $x$--$z$ plane, which are driven by nonlinear mode conversion of the Alfv\'{e}n waves.
The oscillation seems to be uniform in the $z$ direction, which may be a signature of the body waves rather than the surface waves.
The typical amplitude is approximately 10 km s$^{-1}$, which is significantly smaller than the amplitude of the parent Alfv\'{e}n waves.

\begin{figure}
  \includegraphics[scale=0.9]{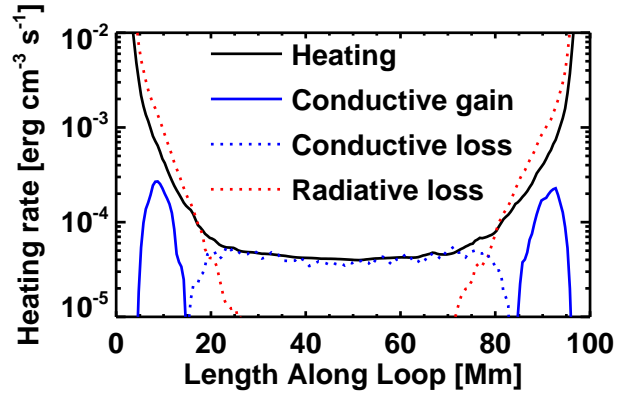}
  \caption{Heating and cooling rate as function of length along loop, averaged over every 1 Mm, $z\in[-1.5,1.5]$ Mm, and $t\in[2,3]$ h.
           The black, red, and blue lines correspond to the heating rate, radiative loss rate, and thermal conduction loss rate, respectively.
           The solid and dotted lines indicate positive and negative values, respectively.}
  \label{fig:ebalance}
\end{figure}

\subsection{Thermal balance}

The thermal structure of the model atmosphere is determined by examining the energy balance between heating and cooling.
As no explicit dissipation terms were included in the basic equations,
the heating was purely derived from the nature of our numerical scheme.
In a previous study, \cite{2014MNRAS.440..971M} developed a means of estimating the numerical heating rates ($\Gamma$),
which was employed here.
$\Gamma$ is the same quantity as $Q_a$ in \cite{2014MNRAS.440..971M} which can be estimated using discretization errors in the equation of total energy.
$\Gamma$ is a good indicator of numerical heating rate per volume especially for the dissipation of Alfv\'{e}n waves, which is 
demonstrated by problems of linear MHD wave dissipation.
The spatial distribution of the heating rate, which is shown in Fig. \ref{fig:ebalance}, decreases exponentially with height below 20 Mm.
The average scale height of the heating rate is approximately 2 Mm in the chromosphere, where the radiative cooling is balanced by the conduction gain and numerical heating.
Further, the heating rate in the corona (20 Mm $<x<$ 80 Mm) is almost spatially uniform. As for the cooling rate, the radiative loss is dominant below 20 Mm, while the thermal conductive loss is dominant in the corona.

\begin{figure}
  \includegraphics[scale=1.0]{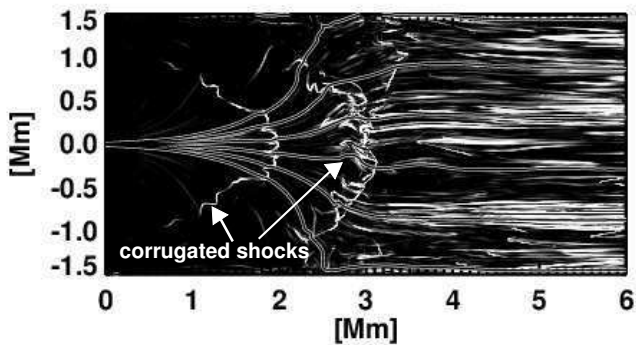}
  \caption{Snapshot of current per unit mass at $t$ = 2.6 h.
           Values less than 10$^{16}$ (in cgs unit) are plotted.
           The solid lines within white envelopes represent the magnetic field lines.
           Two corrugated shocks are indicated by white arrows.}
  \label{fig:crd2}
\end{figure}

The Alfv\'{e}n waves excited randomly at the photosphere form a number of shocklets in the chromosphere.
The shock fronts tend to form a wedge-like structure \citep{1997ApJ...488..854C} determined by the difference in the Alfv\'{e}n
speed across the super-radially expanded flux tube.
As the shocks travel upward, the shock fronts corrugate or fragment into smaller shocks (Fig. \ref{fig:crd2}).
The corrugation process plays an important role in driving the MHD turbulence in the chromosphere.

\begin{figure}
  \includegraphics[scale=1.0]{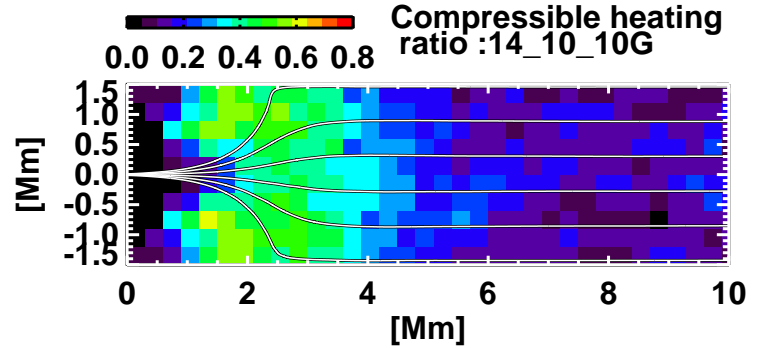}
  \caption{Spatial distribution of compressible vs. total heating amount ratio, averaged over every 150 km in both the $x$ and $z$ directions, and over $t\in[2,3]$ h.
           The white solid lines within black envelopes indicate the magnetic field lines.}
  \label{fig:cmp_heat_ratio}
\end{figure}

The resultant heating rate can be divided into compressible and incompressible heating rate.
We assume that the compressible heating mainly stems from shocks while the incompressible heating originates from 
the direct dissipation of magnetic and velocity shear due to Alfv\'{e}n waves.
The heating rate induced by the compressible process can be roughly estimated as follows.
Let $\Omega$ be a given set of time that has a duration $\tau$ for which the heating rate will be analyzed.
In the present analysis, we will analyze heating rate within time which satisfies $t\in[2,3]$ h.
By using standard manner in the set theory, $\Omega$ can be described by $\Omega=\{t~|~t\in[2,3]{\rm~h}\}$.
Then duration of $\Omega$ is defined by integrating all the members of $\Omega$,
\begin{eqnarray}
  \tau \equiv \int_{\Omega} dt = 1~{\rm h}.
\end{eqnarray}
$\Omega$ can be devided into two subsets.
The first subset of time, $\Omega_C$, is defined as convergent period whose members satisfy $\nabla \cdot \bmath{v}(x,z,t) \le 0$
and can be described as $\Omega_{C}=\{t~|~\nabla \cdot \bmath{v} \le 0\}$.
The other subset of time, $\Omega_D$, is defined as divergent period that can be described as $\Omega_{D}=\{t~|~\nabla \cdot \bmath{v} > 0\}$.
Note that $\Omega_C$ and $\Omega_D$ are the functions of $x$ and $z$.
The duration of each subsets are also defined as,
\begin{eqnarray}
  \tau_C(x,z) &\equiv& \int_{\Omega_C(x,z)} dt, \\
  \tau_D(x,z) &\equiv& \int_{\Omega_D(x,z)} dt.
\end{eqnarray}
The thermal energy supplied by numerical heating rate ($\Gamma$) during $\Omega_{C}$ and $\Omega_{D}$ can be expressed as
\begin{eqnarray}
  E_C'(x,z)&=&\int _{\Omega _C}\Gamma (x,z,t)~dt, \\
  E_D(x,z)&=&\int _{\Omega _D} \Gamma (x,z,t)~dt.
\end{eqnarray}
We assume that the contribution of shear to heating is always active, while the contribution of shocks
is concentrated within periods of time and space where $\nabla \cdot \bmath{v} \le 0$, denoted by $\Omega_C$.
We have then the two periods: 
\begin{eqnarray}
 \Omega_D &{\rm where}&  \Gamma =  \Gamma_{shear} , \\
 \Omega_C &{\rm where}&  \Gamma =  \Gamma_{shear}  + \Gamma_{comp} .
\end{eqnarray}
In order to decompose the contribution of shear and shocks, it is also assumed that $\Gamma_{shear}$ is not varying statistically.
Then the temporal average of $\Gamma_{shear}$ is identical during the two periods:
$\langle \Gamma_{shear} \rangle _C = \langle \Gamma_{shear} \rangle _D$ where
\begin{eqnarray}
 \langle f \rangle _i \equiv \frac{1}{\tau_i} \int _{\Omega_i} f dt
\end{eqnarray}
for $i=C,D$ and $f$ is any functions of time.
Only $\Gamma$ is measured directly, but since $\langle \Gamma_{shear} \rangle_C = \langle \Gamma_{shear} \rangle_D $, we
can write $\langle \Gamma_{comp} \rangle_C $ in terms of $\Gamma$ as:
\begin{eqnarray}
 \langle \Gamma_{comp} \rangle _C = \langle \Gamma \rangle_C - \langle \Gamma \rangle _D
\end{eqnarray}
Using the respective time duration of the two periods, we can finally find
\begin{eqnarray}
  E_C = E_C' - \frac{E_D}{\tau_D} \tau_C,
\end{eqnarray}
which gives the relation between integrated energies.
Although this is a crude approximation, it helps us to distinguish between the compressible and incompressible heating processes.
Fig. \ref{fig:cmp_heat_ratio} shows the spatial distribution of the ratio of $E_C$ to $E_C'+E_D$.
The ratio is averaged over 150 km in both the $x$ and $z$ directions in order to increase the statistics.
The compressible heating rate is almost 50 \% for the 1--4-Mm region, which means that shock heating is effective in this region.

\begin{figure}
  \includegraphics[scale=1.0]{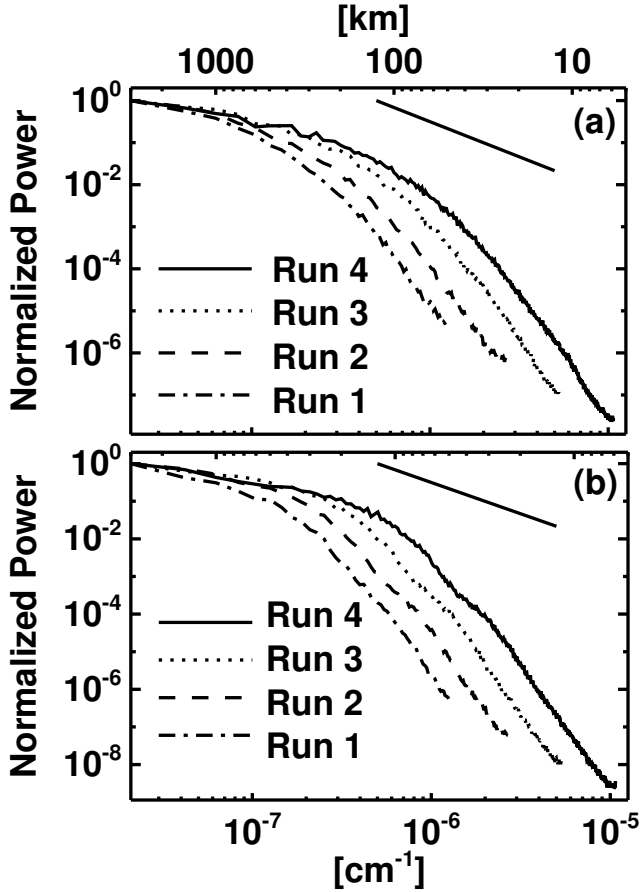}
  \caption{Power spectral density of $B_y$ at (a) 6 Mm and (b) loop top for different spatial resolutions.
           The solid straight line in each panel shows the power law ($\propto k^{-5/3}$) for reference.}
  \label{fig:psd}
\end{figure}

The contribution of the incompressible heating to the total heating rate is 50 \% between 1 and 4 Mm
and more than 90 \% above 4 Mm (Fig. \ref{fig:cmp_heat_ratio}).
The 4-Mm height corresponds to the top of the magnetic canopy, where the flux tube merges with the
neighbouring flux tubes. As the magnetic field strength becomes constant above 4 Mm, the
Alfv\'{e}n speed increases with height because of the density stratification.
Cascading of the Alfv\'{e}n waves across the $z$ direction can be seen in this region.
The cascading process creates a large number of thin current sheets extended toward the $x$ direction.
The wave heating is highly dynamic and occurs on these thin current sheets.

\subsection{Fourier Analysis}

Power spectra usually provide useful information on the properties of turbulent phenomena.
Here, the power spectrum of the magnetic energy in the $z$ direction is estimated using the following definition,
\begin{eqnarray}
  E(x,k) = \frac{1}{2} |\hat{B_y}(x,k)|^2,
\end{eqnarray}
where $\hat{B_y}(x,k)$ is defined by
\begin{eqnarray}
  \hat{B_y}(x,k) \equiv \frac{1}{\sqrt{2\pi}} \int^{W/2}_{-W/2} B_y(x,z) e^{-ikz} dz, \label{eq:fourier}
\end{eqnarray}
and $k$ denotes the wave number in the $z$ direction.
Note that only the $B_y$ is considered, because this component is the dominant energy carrier in our simulation.

Fig. \ref{fig:psd} shows (a) the estimated power spectra at 6 Mm and (b) the loop top for different spatial resolutions.
The horizontal axis on the top of the figure indicates the wavelength, $\lambda=2\pi/k$, in units of kilometre.
Since the initial photospheric disturbance is uniform in $z$ direction, the power at $k>0$ originates from turbulent cascading process.
At all resolutions, the power spectrum decreases with $k$ from the larger scale in the energy injection range toward the smaller dissipation scale.
However, there is no clear dissipation tail which should be exponential-like, instead there is a $k^{-6}$ behavior in the right part of 
wave number range for all runs.
Moreover, there is no clear inertial range in the left part of the spectral range, at least no inertial range common to all runs.
Indeed, for the lowest resolution run, a spectral scaling is not so far from $k^{-5/3}$, but for the 
higher resolution run, the spectrum is flatter than $k^{-1}$ (Fig. \ref{fig:psd}b).
These facts may suggest that the turbulence in the present simulation is not the standard Kolmogorov turbulence.

\begin{figure}
  \includegraphics[scale=1.0]{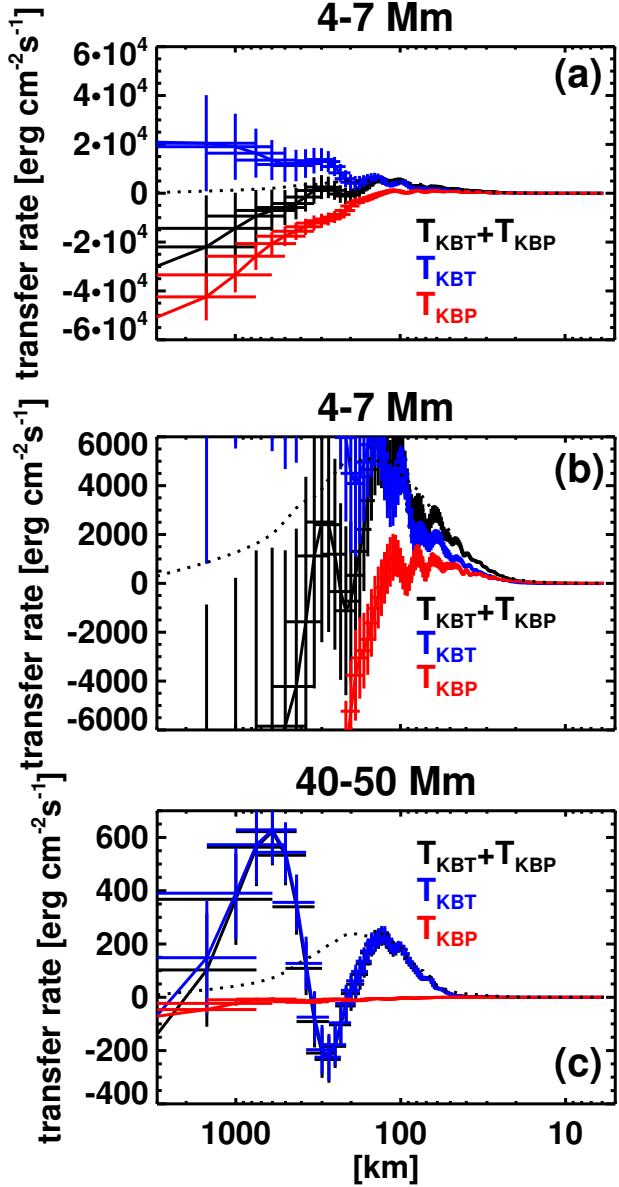}
  \caption{Energy transfer rates for Run 4 as function of wave number in $z$ direction, averaged over $t\in[2,3]$ h (quasi steady-state regime).
           (a) Chromospheric energy transfer rates averaged over $z\in[4,7]$ Mm.
           (b) Identical to (a) but with a smaller range, to emphasize the values around the smaller wave number.
           (c) Energy transfer rates at loop top averaged over $z\in[40,50]$ Mm.
           The black, blue, and red solid lines correspond to 
           $T_{KBT}+T_{KBP}=D_{num}$
           , $T_{KBT}$, and $T_{KBP}$, respectively.
           The dotted lines correspond to 
           $D_{res}=\eta_{num} k^2 |\hat{B}_y(x,k)|^2$.
           The horizontal axis on the top of each panel indicates the wavelength, $\lambda=2\pi/k$, in units of kilometre.}
  \label{fig:transfer}
\end{figure}

Using the induction equation, the temporal evolution of the magnetic power spectra can be derived in the form
\begin{eqnarray}
  \frac{d E(x,k)}{dt} = T_{KBT}(x,k)+T_{KBP}(x,k),
\end{eqnarray}
where the right-hand-side variables are defined as
\begin{eqnarray}
  T_{KBT}(x,k) &=& \hat{B_y^*} \left[\widehat{\bmath{B}\cdot \nabla \bmath{v}|_y}\right] +c.c., \\
  T_{KBP}(x,k) &=& \hat{B_y^*} \left( \left[ \widehat{\bmath{v}\cdot \nabla \bmath{B}|_y} \right]
                                      -\left[ \widehat{B_y \nabla \cdot \bmath{v}} \right] \right) + c.c.,
\end{eqnarray}
where the $~\widehat{.}~$ symbol indicates the finite Fourier transform in a similar manner to Eq. (\ref{eq:fourier}) and
$~c.c.~$ denotes the complex conjugates of the first term on the right-hand side of each equation.
The transfer function, $T_{KBT}$, denotes the energy transfer rate from the kinetic-energy reservoir to
the $k$-component of the magnetic-energy reservoir via fluid motion stretching against the magnetic tension force.
Similarly, $T_{KBP}$ denotes the energy transfer from the kinetic to the magnetic energy through compression against the magnetic pressure.
Note that all the energy transfer processes here arise from the interaction between the $x$ and $z$ components of $\bmath{v}$ and $\bmath{B}$
(the magnetoacoustic waves or the background magnetic field) and the $y$ components of the fluctuations (Alfv\'{e}n waves).
Direct nonlinear energy transfer toward the larger $k$ direction between the Alfv\'{e}n waves is absent from the present 2.5D MHD framework.
In a steady state, the time derivative of $E(x,k)$ is zero and
\begin{eqnarray}
  T_{KBT}+T_{KBP}=0. \label{eq:eqtransfer}
\end{eqnarray}
As the numerical simulation has numerical dissipation, Eq. (\ref{eq:eqtransfer}) is modified to
\begin{eqnarray}
  T_{KBT}+T_{KBP}=D_{num},
\end{eqnarray}
where $D_{num}$ indicates the effect of the numerical dissipation.
Similar analysis of $T_{KBT}$ has been conducted in the context of the solar dynamo \citep{2010ApJ...714.1606P}.

The transfer analysis indicates that the stretching motion against the magnetic tension force makes the largest contribution
to the transfer from the kinetic to the magnetic energy on the small spatial scale in the chromosphere.
Fig. \ref{fig:transfer}a shows the energy transfer rates for Run 4 in the chromosphere, averaged over $x\in[4,7]$ Mm and $t\in[2,3]$ h.
The same quantities are plotted in Fig. \ref{fig:transfer}b, with a smaller energy-transfer-rate range and with the
same wave-number range, in order to emphasize the smaller wave-number values.
The fact that $T_{KBT}>0$ for almost the entire wave-number space suggests that the field stretching transfers energy to the magnetic energy
in each wave-number bin.
The compressible energy transfer, $T_{KBP}$, is negative on the larger scale ($\lambda \gtrsim 100$ km), which means that the divergent motion reduces
the magnetic energy in that case.
On the other hand, the convergent motion supplies the energy on the smaller scale ($\lambda \lesssim 100$ km).

The effective 
Lundquist number $S\equiv l V_{A}/\eta _{num}$ 
is estimated to be 7 $\times$ 10$^4$ in the region of $x\in[4,7]$ 
, where $l$ and $V_A$ indicate a typical length and Alfv\'{e}n speed and $\eta _{num}$ is 
numerical resistivity estimated as follows.
If the numerical dissipation is assumed to be represented exactly by a resistive process having a uniform resistivity $\eta_{num}$,
the transfer rates determined by the resistivity can be written in the form
\begin{eqnarray}
 D_{res} = \eta_{num} k^2|\hat{B_y}(x,k)|^2.
\end{eqnarray}
Using $\eta_{num}$ as a parameter, the high wave number tails ($k>5\times10^{-7}$ cm$^{-1}$) of $D_{num}$ are fitted 
to obtain $\eta_{num}$.
The dotted line in Fig. \ref{fig:transfer}b shows 
$D_{res}$.
Although 
$D_{res}$ and $D_{num}$
have similar curves on the smaller scale, they deviate from each other on the larger scale.
This discrepancy may arise from the poorer statistics available on the larger scale, or from the fact that numerical
dissipation cannot be regarded as a diffusion process on the larger scale.
For the typical length ($l=3$ Mm) and the typical Alfv\'{e}n speed ($V_A=70$ km s$^{-1}$),
$S$ 
is estimated to be $7\times10^4$.
Note that the same approach to estimating 
magnetic Reynolds number
has also been implemented by \cite{2007A&A...476.1113F}
, although their magnetic Reynolds number is neither a magnetic Reynolds number nor a Lundquist number, 
as it is based on the sound speed instead of Alfv\'{e}n speed.

At the loop top ($x\in[40,50]$ Mm), the energy transfer through the stretching motion is dominant over almost the entire wave-number space (Fig \ref{fig:transfer}c).
The contribution from the compressible energy transfer is negative and very small.
The effective 
$S$
is estimated to be $7\times10^5$, if the typical length
($l=3$ Mm) and the typical Alfv\'{e}n speed ($V_A=1700$ km s$^{-1}$) are used.

\section{Discussion}

In this study, a 2.5D MHD simulation for a coronal loop were performed.
It was found that the dissipation of Alfv\'{e}n waves can maintain the hot coronal loop, which satisfies the RTV scaling law,
provided the maximum energy input from the photosphere is available.
The resultant transition region has a multithread structure and exhibits a fluctuating motion similar to the spicule motion.
The model loop consists of numerous thin threads with an isothermal nature and no elemental width.
Although a small spatial grid (from 25 to 3 km) was used in this study, the loop top temperature and density continued to increase with improved spatial resolution.
Both the shock heating and turbulent heating were found to be in effect in the region of 1-4-Mm height, while
the MHD turbulence was found to primarily heat the atmosphere above 4 Mm.
According to the transfer analysis of our MHD simulations, both compressible and field stretching motion
contribute to the turbulent cascade in the upper chromosphere, while only the field stretching motion
is important in the corona.

As the resultant coronal loop obeys RTV theory well, our model has at least two inconsistencies with the recent observations.
First, the obtained temperature distribution is inconsistent with the observed distribution, which has a flatter profile
along the loop \citep{2000ApJ...541.1059A}.
Second, our model cannot explain the over- or under-dense loops that are very common in the sun.
Impulsive heating events such as nanoflares are required in order to explain this behaviour \citep{2006SoPh..234...41K}.
Even though our model corona reveals a 10 \% fluctuation from the average in the heating rate, the system exhibits a quasi-steady state rather than dynamical evolution.

Further, the average height of the transition region in our simulation (7.7 Mm) is larger than the observed height of 2 Mm \citep{1993ApJ...406..319F}.
The important factors necessary to determine the height of the transition region are the heating rate in the corona and the $P_{ram}$ or $P_{mag}$ from the waves.
A larger heating rate in the corona generally produces higher coronal pressure, resulting in a smaller transition-region height from the perspective of the pressure balance between the corona and the chromosphere.
Further, larger wave pressure elevates more material into the chromosphere, thereby generating higher chromospheric pressure, which creates a taller transition region.
In our simulation, both effects create a taller transition region.

The MHD turbulence in the chromosphere may be the key process in the creation of thread-like structures such as spicules along the magnetic field lines.
It was found that the thread-like structures can be produced even if the initial wave perturbations are uniform in the $z$ direction.
This indicates that spicular structures can be created naturally, even when the magnetic patches are uniformly jostled
by the convection and there are no internal flows in the magnetic patches.
This phenomenon occurs when the wave nonlinearity ($\equiv \langle V_y^{2}\rangle^{1/2}/\langle V_A \rangle$)
in the chromosphere is relatively large ($\sim$0.3 in our model), although
further simulations are required in order to confirm the critical wave-nonlinearity values.

The thread-like structures continue to exist in the corona, creating a multi-stranded coronal loop.
The thread width does not have any smallest or elemental size of 300 km,
as was implied by recent observations \citep{2012ApJ...755L..33B,2013ApJ...772L..19B}.
Instead, the width exhibits a power law distribution that creates smaller spatial scales with increased numerical resolution.

In our model, the effects of compressibility, which have been neglected in RMHD, are important to drive the turbulence in the chromosphere.
In the RMHD framework, the $\bmath{v}$ and $\bmath {B}$ disturbances can be expressed using scalar potentials; this approach is based on
the assumption that the wave nonlinearity is small \citep{2011ApJ...736....3V}.
Further, this assumption yields a zero nonlinear term in the 2.5D RMHD framework.
However, the wave nonlinearity is not small ($\sim 0.3$ in our model chromosphere) in some cases, depending on the wave input energy,
which violates the assumption of the RMHD equations.
Instead, the energy cascade can be driven by the nonlinear interactions, which involve compressible processes.

The Alfv\'{e}n waves, or disturbances in the $y$ direction, are primary carriers of the energy flux in our simulation.
This is reasonable, as only Alfv\'{e}n waves are driven at the bottom of simulation box, and
fast and slow waves (disturbances in the $x$--$z$ plane) are then produced by the mode conversion from the Alfv\'{e}n waves.
The dominant wave modes in the energy flux are changed when the different types of photospheric drivers are assumed
\citep{2011ApJ...727...17F,2015ApJ...799....6M}.

Our model only allows shear Alfv\'{e}n waves to exist; however, swirling motions have recently been observed
as magnetic tornadoes \citep{2012Natur.486..505W}, which 
are interpreted as torsional Alfv\'{e}n waves rather than bulk motion, like tornadoes on Earth \citep{2013ApJ...776L...4S}.
Although the nonlinear behaviours of the shear and torsional Alfv\'{e}n waves become identical in the zero-plasma-$\beta$ limit,
a discrepancy appears in the finite-plasma-$\beta$ case.
As the nonlinear steepening of the shear Alfv\'{e}n waves is less effective than that of the torsional waves \citep{2012A&A...544A.127V},
our simulation overestimates the shock heating rate in the chromosphere.

The temperature and density at the loop top increase with the numerical resolution in our simulations.
At first, we expect that the temperature and density should be independent of the resolution if the turbulent
heating is active, or that they may even decrease with the resolution if some other scale-dependent heating acts.
The increase in the temperature and density arises from the increase in the Poynting flux due to the decrease in the numerical dissipation in the chromosphere.
At the loop top, the volumetric heating rate normalized by $B_y^2/2\tau_A$, where $\tau_A=L/V_A$, is almost independent of
the resolution and has a value of approximately 0.3.
This means the typical heating time scale is approximately 3 Alfv\'{e}n transit times.

The turbulence in the corona and that in the chromosphere have different driving mechanisms.
In the corona, the motion against magnetic pressure does not contribute significantly
to the energy transfer in the wave number space 
since $|T_{KBP}| \ll |T_{KBT}|$.
On the other hand, in the chromosphere, 
the compression motion makes non-negligible contributions to the
magnetic-energy transfer from the larger ($\gtrsim 100$ km) to the smaller scale.
In both the corona and the chromosphere, the field stretching motion distributes the energy
throughout the majority of the wave-number space.

The periodic boundary conditions are positioned in the $z$ direction in our simulation, 
and one important problem arising from the boundary condition must be pointed out.
The resonant absorption of Alfv\'{e}n waves, which is considered to be very effective 
in the coronal loop \citep{1978ApJ...226..650I}, may be neglected, as our model does not have density gradients at the loop boundary.
However, our model does reveal numerous thin threads that have density gradients inside the loop.
This phenomenon could cause a self-consistent resonant absorption process  \citep{1998ApJ...493..474O}, although 
further analysis is required in order to confirm the contribution from this process.

\section{Conclusion}

The coronal structure was reproduced as a natural consequence of Alfv\'{e}n wave injection from the photosphere using 2.5D MHD simulations.
The resultant coronal loop reached a quasi-steady state that obeyed the RTV scaling law.
It was found that both the shock heating rate and the turbulent heating rate are almost comparable below the magnetic canopy (4 Mm).
The shock heating mechanism was dominated by the turbulent heating mechanism above the magnetic canopy.
The coronal structure was maintained even when an unprecedentedly high spatial resolution of $3$ km was applied,
although the resultant temperature and density continued to increase with the resolution.
Although the competition between the shock and turbulent heating was investigated, 
our model might underestimate the effect of the other important heating processes, such as the resonant absorption.
Further study is necessary in order to distinguish between such processes.

Although our 2.5D model exhibited a variety of properties compared to the previous 1D models, several important
features were omitted because of the 2.5D approximation.
One may speculate that a full 3D treatment would have a non-negligible impact on the MHD turbulence and the magnetic reconnection.
Although the finite wave nonlinearity facilitates MHD turbulence driving, even in 2.5 dimensions,
the interactions achieved in that case may be smaller than those obtainable using nonlinear terms in a full 3D RMHD formulation \citep{1990JGR....9510291Z}.
Further, it would be possible to derive other important 3D effects from the magnetic reconnection determined via a full 3D treatment.
In a 3D configuration, magnetic reconnections could occur at the locations of the thin current sheets.
Note that magnetic reconnection can act as an additional heating source by generating secondary MHD waves \citep{2010PASJ...62..993K}.

The present study presents the first direct MHD simulation able to distinguish 
between the shock and turbulent heating mechanisms in a single coronal loop.
The different heating mechanisms lead to different resultant coronal temperature and density profiles.
Since the coronal temperature and density determine the mass loss from the stellar objects,
the coronal heating theory has significant impacts on the prospects not only of the present solar wind theory but also of 
the mass loss theory from the young sun \citep{2013PASJ...65...98S} or the other stellar objects \citep{2011ApJ...741...54C}.
To fully understand the response of the solar atmosphere from the photospheric perturbation
will require a wide parameter survey using the model developed here.
Also, we have to fill a gap between heating the closed corona and the open corona, since
waves don't behave in the same way in the two regions.

\section*{Acknowledgments}

The author appreciate the anonymous referee for constructive comments.
The numerical computations were conducted on a Cray XC30 supercomputer at the Centre for Computational Astrophysics,
National Astronomical Observatory of Japan.
This work was supported by JSPS KAKENHI, Grant Number 50728326.

\bibliographystyle{mn2e}


\end{document}